\documentclass[a4paper,11pt]{article}
\usepackage{pos_mod}

\usepackage{tabularx}
\usepackage{longtable} 
\usepackage{multicol}
\usepackage{multirow}

\renewcommand{\arraystretch}{1.2}

\newcommand{\ra}[1]{\renewcommand{\arraystretch}{#1}}
\newcommand{\rb}[1]{\renewcommand{\tabcolsep}{#1}}

\title{Implications of LHCb Data for Lepton Flavour Universality Violation}

\author[a]{T.~Hurth}
\author[b,c]{F.~Mahmoudi}
\author[d]{D.~Mart\'inez Santos}
\author[e]{S.~Neshatpour}

\affiliation[a]{PRISMA+ Cluster of Excellence and  Institute for Physics (THEP),\\
Johannes Gutenberg University, D-55099 Mainz, Germany}

\affiliation[b]{Universit\'e de Lyon, Universit\'e Claude Bernard Lyon 1, CNRS/IN2P3,\\
Institut de Physique des 2 Infinis de Lyon, UMR 5822, F-69622, Villeurbanne, France}

\affiliation[c]{Theoretical Physics Department, CERN,
CH-1211 Geneva 23, Switzerland}

\affiliation[d]{Instituto Galego de F\'isica de Altas Enerx\'ias,
Universidade de Santiago de Compostela, Spain}

\affiliation[e]{INFN-Sezione di Napoli,
Via Cintia, 80126 Napoli, Italia}

\emailAdd{Tobias.Hurth@cern.ch,nazila@cern.ch,Diego.Martinez.Santos@cern.ch,\\ neshatpour@na.infn.it\\
CERN-TH-2021-180, MITP-21-053}

\abstract{We analyse the new physics implications of theoretically clean $b \to s$ observables in a model-independent approach and compare their coherence with the implications of other rare $B$-decays. A~statistical comparison is done between the New Physics explanation and hadronic contributions as the source of the anomalies in angular observables of the $B\to K^*\mu\mu$ decay. 
We make projections for future measurements that indicate that LHCb will be in the position to discover lepton non-universality via a single observable using the Run 3 data. 
The global fit of rare $B$-decays is given within a multidimensional fit involving all the 20 relevant Wilson coefficients.}

\begin{document}
\maketitle

\section{Theoretically clean vs the rest of the observables}
Recent LHCb measurements have indicated tensions with the Standard Model (SM) predictions in a number of $b \to s$ decays.  
There are tensions in the angular observables of the $B \to K^* \mu^+\mu^-$ decay with the most significant tension in the $P_5^{\prime}$ observable~\cite{LHCb:2020lmf}. Similar tensions have also been measured in the $B^+ \to K^{*+}\mu^+\mu^-$ decay~\cite{LHCb:2020gog}.
Furthermore, the branching ratio of several $B$-decays such as $B\to K \mu^+\mu^-, B_s\to \phi \mu^+\mu^-$ and $\Lambda_b \to \Lambda \mu^+\mu^-$ have been measured to be below the SM prediction~\cite{LHCb:2014cxe,LHCb:2015tgy,LHCb:2021zwz}.
The recent LHCb measurement on the lepton flavour universality violating (LFUV) observable $R_K$ has confirmed the tension with the SM with $3.1\sigma$ significance~\cite{LHCb:2021trn}. LHCb has measured similar deviations in $R_{K^*}$ in the two low $q^2$ bins with $2.3$ and $2.5\sigma$ significance~\cite{LHCb:2017avl}.

To study the New Physics (NP) implication of these measurements, all the relevant $B$-decay observables should be considered.
However, the precision of the theoretical predictions is not the same for all these observables. 
Due to the cancellation of hadronic uncertainties in the numerator and the denominator,
the LFUV observables $R_{K^{(*)}}={\rm BR}(B\to K^{(*)}\mu\mu)/{\rm BR}(B\to K^{(*)}ee)$ are predicted very precisely in the SM, with theoretical uncertainty less than 1~(3)\% for the $q^2~\in~[1.1,6] ([0.045,1.1])$ GeV$^2$ bin.  Another clean observable with small theoretical uncertainty (less than 5\%) is the branching ratio of the $B_s \to \mu^+ \mu^-$ decay. On the other hand, the rest of the $b \to s$ observables in general suffer from larger theoretical uncertainties due to hadronic contributions.  Although with the appropriate choice of angular observables, less sensitivity from local form factor uncertainties is achievable, there are still contributions from power corrections of non-local hadronic effects which are not well-known within QCD factorisation and are usually ``guesstimated'' (for a study of the impact of the local and non-local hadronic uncertainties on NP fits see Ref.~\cite{Hurth:2016fbr}).

Therefore, we separate the theoretically ``clean observables''
from the rest of the $b \to s$ observables and compare the NP implications 
and coherence of NP fits to these two data sets.
For the analysis we have used the {\ttfamily SuperIso} public program~\cite{Mahmoudi:2007vz}. 
From Table~\ref{tab:Clean_vs_rest_1D}, we see that in the one-dimensional fits to the clean observables there are several NP scenarios explaining the data with more than $4\sigma$ significance better than the SM~\cite{Hurth:2021nsi}. 
\begin{table}[bh!]
\begin{center}
\setlength\extrarowheight{0pt}
\scalebox{0.75}{
\begin{tabular}{|l|r|r|c|}
\hline 
\multicolumn{4}{|c|}{\footnotesize \qquad Only $R_{K^{(*)}}, B_{s,d} \to \mu^+ \mu^-$ \qquad ($\chi^2_{\rm SM}=	28.19	$)}\\ \hline	 									
& b.f. value & $\chi^2_{\rm min}$ & ${\rm Pull}_{\rm SM}$  \\										
\hline \hline										
$\delta C_{9} $    	& $ 	-1.00	\pm	6.00	 $ & $ 	28.1	 $ & $	0.2	\sigma	 $  \\[-4pt]
$\delta C_{9}^{e} $    	& $ 	0.80	\pm	0.21	 $ & $ 	11.2	 $ & $	4.1	\sigma	 $  \\[-2pt]
$\delta C_{9}^{\mu} $    	& $ 	-0.77	\pm	0.21	 $ & $ 	11.9	 $ & $	4.0	\sigma	 $  \\
\hline										
$\delta C_{10} $    	& $ 	0.43	\pm	0.24	 $ & $ 	24.6	 $ & $	1.9	\sigma	 $  \\[-4pt]
$\delta C_{10}^{e} $    	& $ 	-0.78	\pm	0.20	 $ & $ 	9.5	 $ & $	4.3	\sigma	 $  \\[-2pt]
$\delta C_{10}^{\mu} $    	& $ 	0.64	\pm	0.15	 $ & $ 	7.3	 $ & $	4.6	\sigma	 $  \\
\hline							          			
$\delta C_{\rm LL}^e$	& $ 	0.41	\pm	0.11	 $ & $ 	10.3	 $ & $	4.2	\sigma	 $  \\[-4pt]
$\delta C_{\rm LL}^\mu$	& $ 	-0.38	\pm	0.09	 $ & $ 	7.1	 $ & $	4.6	\sigma	 $  \\
\hline										
\end{tabular} \qquad
\begin{tabular}{|l|r|r|c|}
\hline 
\multicolumn{4}{|c|}{\footnotesize All obs. except $R_{K^{(*)}}, B_{s,d}\to\mu^+ \mu^-$ \; ($\chi^2_{\rm SM}=	200.1	$)}\\ \hline	 									
& b.f. value & $\chi^2_{\rm min}$ & ${\rm Pull}_{\rm SM}$  \\	 									
\hline \hline										
$\delta C_{9} $    	& $ 	-1.01	\pm	0.13	 $ & $ 	158.2	 $ & $	6.5	\sigma	 $  \\[-4pt]
$\delta C_{9}^{e} $    	& $ 	0.70	\pm	0.60	 $ & $ 	198.8	 $ & $	1.1	\sigma	 $  \\[-2pt]
$\delta C_{9}^{\mu} $    	& $ 	-1.03	\pm	0.13	 $ & $ 	156.0	 $ & $	6.6	\sigma	 $  \\
\hline										
$\delta C_{10} $    	& $ 	0.34	\pm	0.23	 $ & $ 	197.7	 $ & $	1.5	\sigma	 $  \\[-4pt]
$\delta C_{10}^{e} $    	& $ 	-0.50	\pm	0.50	 $ & $ 	199.0	 $ & $	1.0	\sigma	 $  \\[-2pt]
$\delta C_{10}^{\mu} $    	& $ 	0.41	\pm	0.23	 $ & $ 	196.5	 $ & $	1.9	\sigma	 $  \\
\hline							          			
$\delta C_{\rm LL}^e$	& $ 	0.33	\pm	0.29	 $ & $ 	198.9	 $ & $	1.1	\sigma	 $  \\[-4pt]
$\delta C_{\rm LL}^\mu$	& $ 	-0.75	\pm	0.13	 $ & $ 	167.9	 $ & $	5.7	\sigma	 $  \\
\hline										
\end{tabular}
}
\caption{\small Comparison of one operator NP fits to clean observables on the left and to the rest of the $b\to s$ observables on the right (assuming 10\% error for the power corrections).
\label{tab:Clean_vs_rest_1D}} 
\end{center} 
\end{table}
\vspace{-0.2cm}
For the one-dimensional fit to  all $b \to s$ observables except the clean ones (right panel of Table.~\ref{tab:Clean_vs_rest_1D}), the most favoured scenario is NP in $C_9^{(\mu)}$ with a significance of ~$6.5\sigma$. However, this significance depends on the choice of form factors as well as the guesstimated size of the non-factorisable power corrections (here assumed to be 10\% compared to leading order QCD factorisation contributions).
Compared with the NP fit to the rest of the observables, there are favoured scenarios such as NP in $C_9^\mu$, resulting in coherent best fit values for both sets of observables.   
This is also the most favoured scenario in the global fit where the clean observables and the rest of the $b\to s$ observables are considered together~\cite{Hurth:2021nsi} (see Refs.~\cite{Geng:2021nhg,Altmannshofer:2021qrr,Alguero:2021anc} for other recent global fits).

\section{NP or hadronic contributions in $B\to K^* \mu \mu$ observables}
The impact of the guesstimated size of power corrections on the significance of NP in $C_9$ can be clearly seen by describing the $B \to K^* \mu^+\mu^-$ decay in terms of helicity amplitudes, with NP effects in $C_9$ (and $C_7$) and power corrections $h_\lambda$, both contributing to the vectorial helicity amplitude~\cite{Jager:2012uw}
\begin{align}
   H_V(\lambda) &=-i\, N^\prime \Big\{ C_9^{\rm eff} \tilde{V}_{\lambda} - C_{9}'  \tilde{V}_{-\lambda}
      + \frac{m_B^2}{q^2} \Big[\frac{2\,\hat m_b}{m_B} (C_{7}^{\rm eff} \tilde{T}_{\lambda} - C_{7}'  \tilde{T}_{-\lambda})
      - 16 \pi^2 \left(\text{LO QCDf}+h_\lambda\right) \Big] \Big\} \,.
\end{align}
Instead of making assumptions on the size of the power corrections, these contributions can be parameterised by a number of free parameters and fitted directly to the data.
A general description of the power corrections involves several free parameters~\cite{Ciuchini:2015qxb,Chobanova:2017ghn} which with the current experimental data results in fitted parameters that are loosely constrained~\cite{Hurth:2020rzx}.
A minimalistic description of the hadronic effect is given by~\cite{Hurth:2020rzx,Neshatpour:2017qvi}
\begin{align}
 h_\lambda (q^2)= -\frac{\tilde{V}_\lambda(q^2)}{16 \pi^2} \frac{q^2}{m_B^2}  \Delta C_9^{\lambda,\rm{PC}}\,,
\end{align}
which involves only three real free parameters corresponding to each helicity $\lambda=0,\pm$ (six if assumed complex). This description with smaller degrees of freedom (dof) in principle has a better chance of giving a constrained fit and can be considered as a null test for NP; if the three fitted hadronic parameters (one free parameter corresponding to each helicity) differ from each other, NP in $\delta C_9^{\rm NP}$ can be ruled out. Although it is possible that the fitted power corrections for each helicity are very similar to mimic NP in $\delta C_9^{\rm NP}$, it is highly improbable, furthermore there are theoretical arguments that the positive helicity amplitude should be suppressed compared to the two other helicities~\cite{Jager:2014rwa}.

\begin{table}[b!]
\ra{1.}
\rb{1.3mm}
\begin{center}
\setlength\extrarowheight{2pt}
\scalebox{0.85}{
\begin{tabular}{|c|c|c|}
\hline
 \multicolumn{3}{|c|}{Best fit values of hadronic power corrections}\\ \hline  \hline
$\Delta C_9^{+,{\rm PC}}$ & $\Delta C_9^{-,{\rm PC}}$ & $\Delta C_9^{0,{\rm PC}}$ \\  
\hline
$  5.43	\pm	6.22 $ & $ -1.06	\pm	0.21 $  & $ -0.73	\pm	0.52 $  \\  
\hline
\end{tabular}  \qquad 
\begin{tabular}{|l|c|c|}
\hline
\multicolumn{3}{|c|}{Significance of NP and hadronic p.c. fits}           \\  
\hline \hline
nr. of dof & $\footnotesize 1\; (\delta C_9^{\rm NP})$ & $\footnotesize 3\; (\Delta C_9^{\lambda,{\rm PC}})$ \\ [1pt]
\hline
0 (plain SM)					 	&	 $6.0\sigma$	  	&	 $5.4\sigma$ \\
1 {\small(Real $\delta C_9$)}		&	 $\text{---}$	  	&	 $0.6\sigma$\\ 
\hline
\end{tabular} 
}
\caption{\small
On the left, fit of hadronic power corrections for the three helicities ($\lambda=0,\pm$) with real $\Delta C_9^{\lambda,{\rm PC}}$, using the data on $B\to K^* \bar\mu\mu/\gamma$ observables with $q^2$ bins $\leqslant 8\text{ GeV}^2$. On the right, the significance of the improved description of the hadronic fit as well as the NP fit compared to the SM and to each other.
\label{tab:DeltaC9pc}
}
\end{center} 
\end{table}  
For the fit do data, we consider only the experimental measurements on $B \to K^* \mu^+ \mu^-$ observables in the $q^2 \leq 8$ GeV$^2$ bins  since the power corrections for the low- and high-$q^2$ regions are not necessarily the same.
From the left panel of Table~\ref{tab:DeltaC9pc}, it is clear that although the central value of the best fit point for each helicity is different, the three free parameters cannot be strongly constrained and are compatible with each other within 68\% confidence interval. 
As given in the right panel of Table~\ref{tab:DeltaC9pc}, including either NP contributions ($\delta C_9^{\rm NP}$) or power corrections ($\Delta C_9^{\lambda,{\rm NP}}$), a better description of the data is obtained with a significance of more than $5\sigma$ compared to the SM.    
It should be noted that the NP scenario with $\delta C_9^{\rm NP}$ contributions is embedded in the hadronic fit, hence it is possible to make a statistical comparison between the two fits. And as given in the right panel of Table~\ref{tab:DeltaC9pc}, the improvement of the hadronic fit compared to the NP description is less than $1\sigma$ suggesting that there is no indication to introduce two more dof for the hadronic fit\footnote{Assigning the global $\delta C_9$ as a nuisance parameter to take into account unknown power corrections -- as done for example in Ref.~\cite{Isidori:2021vtc} --  is inappropriate  as there is no theory indication that the three helicities would be described by a common hadronic effect. Even considering the weak sensitivity on the positive helicity, at least two independent free parameters would be necessary to describe the power corrections.}.

\section{Future projections of clean observables}
We consider three benchmark points for the planned LHCb upgrades and make predictions for the clean observables. For the benchmarks, we consider the two LHCb upgrades with 50 and 300~fb$^{-1}$ integrated luminosity as well as an intermediate stage with 18 fb$^{-1}$ of data. Assuming that in future measurements, the current experimental central values remain the same, with the future reduced experimental uncertainties (see~\cite{Hurth:2021nsi} for details) it is not possible to get acceptable fits. 
\begin{table}[t!]
\begin{center}
\setlength\extrarowheight{3pt}
\scalebox{0.85}{
\begin{tabular}{|c||c|c|c|}
\hline 
  \multicolumn{4}{|c|}{Pull$_{\rm SM}$ with $R_{K^{(*)}}$ and ${\rm BR}(B_s\to \mu^+ \mu^-)$ prospects} \\ [-2pt]
\hline 
 LHCb lum.		 &  18 fb$^{-1}$ & 50 fb$^{-1}$   &   300 fb$^{-1}$  \\[-4pt]
\hline
$\delta C_{9}^{\mu}$	 	 & $ 6.5\sigma	 $ & $ 14.7\sigma	 $ &  $ 21.9\sigma	 $   \\[-6pt]
$\delta C_{10}^{\mu}$	 	 & $ 7.1\sigma	 $ & $ 16.6\sigma	 $ &  $ 25.1\sigma	 $   \\[-6pt]
$\delta C_{LL}^{\mu}$	 	 & $ 7.5\sigma	 $ & $ 17.7\sigma	 $ &  $ 26.6\sigma	 $   \\
\hline
\end{tabular}
}\vspace*{0.1cm}
\caption{\small
Predictions of Pull$_{\rm SM}$ for the LHCb upgrade scenarios with 18, 50 and 300 fb$^{-1}$ luminosity collected, for the fit to $\delta C_9^\mu$, $\delta C_{10}^\mu$ and $\delta C_{LL}^\mu$ (as given in the left panel of Table~\ref{tab:Clean_vs_rest_1D}).
\label{tab:LFUV_Bsmumu_projections}}
\end{center} 
\end{table}
\begin{figure}[htb]
\begin{center}
\includegraphics[width=0.45\textwidth]{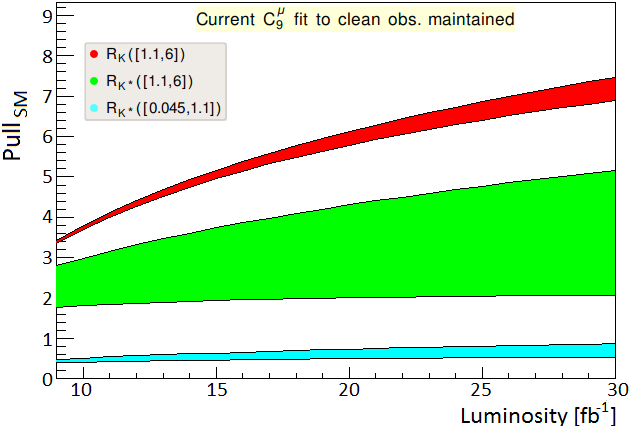}
\includegraphics[width=0.45\textwidth]{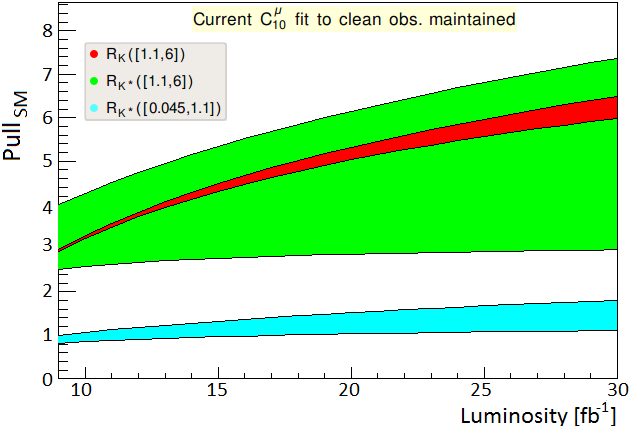}
\caption{\small
Significance of Pull$_{\rm SM}$ for each of the projected LFUV observables, individually.
\label{fig:projections}}
\end{center}
\end{figure}

Instead, we make an equally strong assumption;  we presume that future data correspond to projecting the observables with the current fitted values of each of the three most favoured scenarios of the left panel in Table~\ref{tab:Clean_vs_rest_1D}. As given in Table~\ref{tab:LFUV_Bsmumu_projections}, already with 18 fb$^{-1}$ data, the NP significance will be more than $6\sigma$ in all three scenarios.
However, the significance is quite dependent on the presumed reduction in statistical uncertainties, as can be seen in Fig.~\ref{fig:projections} where Pull$_{\rm SM}$ is  shown for each of the individual LFUV observables when assuming the current central value of $C_9^\mu$ ($C_{10}^\mu$) from the clean observables remains unchanged.
The lower [upper] limit in each band is when assuming current systematic uncertainties do not improve [having ultimate systematic uncertainty of 1\% for the LFUV observables and 5\% for BR($B_s \to \mu^+ \mu^-$)]. 
For the $C_9^\mu$ ($C_{10}^\mu$) scenario, $R_K$ alone can reach $5\sigma$ significance with $\sim 15$ ($20$) fb$^{-1}$ integrated luminosity.

\section{Multidimensional global fit and look-elsewhere effect}
NP does not necessarily present itself in only one or two operator structures, and in principle all the 20 relevant Wilson coefficients can receive NP contributions. 
Furthermore, while look-elsewhere effect (LEE) can be introduced when focusing on a subset of observables,  this can also happen when choosing a posteriori one and/or two  operators.
However, in the case where the fit includes all relevant observables and the
maximum number of Wilson coefficients are set to be free, then LEE is avoided
as there are no a posteriori decisions and the p-values take into account the
number of degrees of freedom and finally insensitive parameters and flat directions can be eliminated based on profile likelihoods and correlations of the fit.

\begin{table}[h!]
\begin{center}
\setlength\extrarowheight{0pt}
\scalebox{0.7}{
\begin{tabular}{|c|c|c|c|}
\hline																
\multicolumn{4}{|c|}{All observables  with $\chi^2_{\rm SM}=	 	225.8			$} \\											
\multicolumn{4}{|c|}{$\chi^2_{\rm min}=	 	151.6	;\quad {\rm Pull}_{\rm SM}=	5.5 (5.6)	\sigma$} \\											
\hline \hline																
\multicolumn{2}{|c|}{$\delta C_7$} &  \multicolumn{2}{c|}{$\delta C_8$}\\																
\multicolumn{2}{|c|}{$	0.05	\pm	0.03	$} & \multicolumn{2}{c|}{$	-0.70	\pm	0.40	$}\\								
\hline																
\multicolumn{2}{|c|}{$\delta C_7^\prime$} &  \multicolumn{2}{c|}{$\delta C_8^\prime$}\\																
\multicolumn{2}{|c|}{$	-0.01	\pm	0.02	$} & \multicolumn{2}{c|}{$	0.00	\pm	0.80	$}\\								
\hline																
$\delta C_{9}^{\mu}$ & $\delta C_{9}^{e}$ & $\delta C_{10}^{\mu}$ & $\delta C_{10}^{e}$ \\																
$	-1.16	\pm	0.17	$ & $	-6.70	\pm	1.20	$ & $	0.20	\pm	0.21	$ & degenerate w/ $C_{10}^{\prime e}$ \\
\hline\hline																
$\delta C_{9}^{\prime \mu}$ & $\delta C_{9}^{\prime e}$ & $\delta C_{10}^{\prime \mu}$ & $\delta C_{10}^{\prime e}$ \\																
$	0.09	\pm	0.34	$ & $	1.90	\pm	1.50	$ & $	-0.12	\pm	0.20	$ & degenerate w/ $C_{10}^{ e}$ \\
\hline\hline																
$C_{Q_{1}}^{\mu}$ & $C_{Q_{1}}^{e}$ & $C_{Q_{2}}^{\mu}$ & $C_{Q_{2}}^{e}$ \\																
$	0.04	\pm	0.10	$ & $	-1.50	\pm	1.50	$ & $	-0.09	\pm	0.10	$ & $	-4.10	\pm	1.5	$ \\
\hline\hline																
$C_{Q_{1}}^{\prime \mu}$ & $C_{Q_{1}}^{\prime e}$ & $C_{Q_{2}}^{\prime \mu}$ & $C_{Q_{2}}^{\prime e}$ \\																
$	0.15	\pm	0.10	$ & $	-1.70	\pm	1.20	$ & $	-0.14	\pm	0.11	$ & $	-4.20	\pm	1.2	$ \\
\hline																
\end{tabular}
} 
\caption{\small 20-dimensional global fit to the $b \to s$ data, assuming 10\% error for the power corrections.
\label{tab:ALL_20D_C78910C12primes}} 
\end{center} 
\end{table}
\vspace{-0.5cm}
In Table.~\ref{tab:ALL_20D_C78910C12primes} we present the 20-dimensional global fit where we obtain Pull$_{\rm SM}=5.5\sigma$. However, considering that two of the Wilson coefficients are degenerate and taking into account the criterion presented in Refs.~\cite{Arbey:2018ics,Hurth:2018kcq}, the effective degrees of freedom are 19 resulting in Pull$_{\rm SM}=5.6\sigma$.

\section{Conclusions}
The $R_K$ and $R_{K^{*}}$ ratios measured by the LHCb collaboration suggest lepton flavour universality violating new physics.
This implication is enforced by considering the rest of the $b\to s$ observables. However, some of the latter observables might suffer from  underestimated non-local hadronic uncertainties. 
We suggested a minimal description of these contributions which can work as a null test for new physics. Nonetheless, with the current data no conclusive judgment is possible.
Moreover, we showed that assuming any of the favoured new physics scenarios remain, future LHCb measurements of lepton flavour universality violating observables can establish beyond the Standard Model physics with more than 5$\sigma$ significance already with 18~fb$^{-1}$ data.
Furthermore, for an unbiased determination of the new physics structure, we also considered a 20-dimensional fit, still finding a large significance for the new physics description of the $b\to s$ data.

\vspace{-0.1cm}


\begin{thebibliography}{99}
\footnotesize



\bibitem{LHCb:2020lmf}
R.~Aaij \textit{et al.} [LHCb],
Phys. Rev. Lett. \textbf{125} (2020) no.1, 011802
[arXiv:2003.04831].


\bibitem{LHCb:2020gog}
R.~Aaij \textit{et al.} [LHCb],
Phys. Rev. Lett. \textbf{126} (2021) no.16, 161802
[arXiv:2012.13241].


\bibitem{LHCb:2014cxe}
R.~Aaij \textit{et al.} [LHCb],
JHEP \textbf{06} (2014), 133
[arXiv:1403.8044].

\bibitem{LHCb:2015tgy}
R.~Aaij \textit{et al.} [LHCb],
JHEP \textbf{06} (2015), 115
[erratum: JHEP \textbf{09} (2018), 145]
[arXiv:1503.07138].


\bibitem{LHCb:2021zwz}
R.~Aaij \textit{et al.} [LHCb],
[arXiv:2105.14007].


%

\bibitem{LHCb:2021trn}
R.~Aaij \textit{et al.} [LHCb],
[arXiv:2103.11769].



\bibitem{LHCb:2017avl}
R.~Aaij \textit{et al.} [LHCb],
JHEP \textbf{08} (2017), 055
[arXiv:1705.05802].

\bibitem{Hurth:2016fbr}
T.~Hurth, F.~Mahmoudi and S.~Neshatpour,
Nucl. Phys. B \textbf{909}, 737-777 (2016)
[arXiv:1603.00865].

 \bibitem{Mahmoudi:2007vz}
  F.~Mahmoudi,
  Comput.\ Phys.\ Commun.\  {\bf 178} (2008) 745, 
[arXiv:0710.2067]; 
F.~Mahmoudi,
  Comput.\ Phys.\ Commun.\  {\bf 180} (2009) 1579, 
[arXiv:0808.3144]; 
 F.~Mahmoudi,
  Comput.\ Phys.\ Commun.\  {\bf 180} (2009) 1718;
%
S.~Neshatpour and F.~Mahmoudi, PoS TOOLS2020 (2021) 036, 
[arXiv:2105.03428].


\bibitem{Hurth:2021nsi}
T.~Hurth, F.~Mahmoudi, D.~M.~Santos and S.~Neshatpour,
[arXiv:2104.10058].



\bibitem{Geng:2021nhg}
L.~S.~Geng, B.~Grinstein, S.~J\"ager, S.~Y.~Li, J.~Martin Camalich and R.~X.~Shi,
[arXiv:2103.12738].

\bibitem{Altmannshofer:2021qrr}
W.~Altmannshofer and P.~Stangl,
[arXiv:2103.13370].

\bibitem{Alguero:2021anc}
M.~Alguer\'o, B.~Capdevila, S.~Descotes-Genon, J.~Matias and M.~Novoa-Brunet,
[arXiv:2104.08921].


\bibitem{Jager:2012uw}
S.~J\"ager and J.~Martin Camalich,
JHEP \textbf{05} (2013), 043
[arXiv:1212.2263].


\bibitem{Ciuchini:2015qxb}
M.~Ciuchini, M.~Fedele, E.~Franco, S.~Mishima, A.~Paul, L.~Silvestrini and M.~Valli,
JHEP \textbf{06} (2016), 116
[arXiv:1512.07157].

\bibitem{Chobanova:2017ghn}
V.~G.~Chobanova, T.~Hurth, F.~Mahmoudi, D.~Martinez Santos and S.~Neshatpour,
JHEP \textbf{07} (2017), 025
[arXiv:1702.02234].


\bibitem{Hurth:2020rzx}
T.~Hurth, F.~Mahmoudi and S.~Neshatpour,
Phys. Rev. D \textbf{102} (2020) no.5, 055001
[arXiv:2006.04213].


\bibitem{Neshatpour:2017qvi}
S.~Neshatpour, V.~G.~Chobanova, T.~Hurth, F.~Mahmoudi and D.~Martinez Santos,
Proceedings of 52nd Rencontres de Moriond on QCD and High Energy Interactions, pp. 87–90, 2017
[arXiv:1705.10730].

\bibitem{Jager:2014rwa}
S.~J\"ager and J.~Martin Camalich,
Phys. Rev. D \textbf{93} (2016) no.1, 014028
[arXiv:1412.3183].



\bibitem{Isidori:2021vtc}
G.~Isidori, D.~Lancierini, P.~Owen and N.~Serra,
Phys. Lett. B \textbf{822} (2021), 136644
[arXiv:2104.05631].





\bibitem{Arbey:2018ics}
A.~Arbey, T.~Hurth, F.~Mahmoudi and S.~Neshatpour,
Phys. Rev. D \textbf{98} (2018) no.9, 095027
[arXiv:1806.02791].




\bibitem{Hurth:2018kcq}
T.~Hurth, A.~Arbey, F.~Mahmoudi and S.~Neshatpour,
Nucl. Part. Phys. Proc. \textbf{303-305} (2018), 2-7
[arXiv:1812.07602].

\end{thebibliography}
\end{document}